# Profit over Proxies: A Scalable Bayesian Decision Framework for Optimizing Multi-Variant Online Experiments

Srijesh Pillai
Department of Computer Science & Engineering
Manipal Academy of Higher Education
Dubai, UAE
srijesh.nellaiappan@dxb.manipal.edu

Rajesh Kumar Chandrawat
Department of Mathematics
Manipal Academy of Higher Education
Dubai, UAE
rajesh.chandrawat@manipaldubai.com

*Abstract* — Online controlled experiments (A/B tests) are fundamental to data-driven decision-making in the digital economy. However, their real-world application is frequently compromised by two critical shortcomings: the use of statistically flawed heuristics like "p-value peeking", which inflates false positive rates, and an over-reliance on proxy metrics like conversion rates, which can lead to decisions that inadvertently harm core business profitability. This paper addresses these challenges by introducing a comprehensive and scalable Bayesian decision framework designed for profit optimization in multi-variant (A/B/n) experiments.

We propose a hierarchical Bayesian model that simultaneously estimates the probability of conversion (using a Beta-Bernoulli model) and the monetary value of that conversion (using a robust Bayesian model for the mean transaction value). Building on this, we employ a decision-theoretic stopping rule based on Expected Loss, enabling experiments to be concluded not only when a superior variant is identified but also when it becomes clear that no variant offers a practically significant improvement (stopping for futility). The framework successfully navigates "revenue traps" where a variant with a higher conversion rate would have resulted in a net financial loss, correctly terminates futile experiments early to conserve resources, and maintains strict statistical integrity throughout the monitoring process.

Ultimately, this work provides a practical and principled methodology for organizations to move beyond simple A/B testing towards a mature, profit-driven experimentation culture, ensuring that statistical conclusions translate directly to strategic business value.

## I. Introduction

In the contemporary digital economy, the capacity for rapid learning and iterative improvement is a primary driver of competitive advantage. At the heart of this capability lies the practice of online controlled experimentation, commonly known as A/B testing. From e-commerce giants optimizing checkout funnels to technology firms refining user interfaces, these experiments form the bedrock of a data-driven culture, providing a rigorous methodology for validating hypotheses and quantifying the impact of change [1]. The insights gleaned from this process are the engine of innovation, guiding product development, marketing strategies, and user experience enhancements that can yield substantial returns.

Despite its foundational importance, the real-world application of online experimentation is often plagued by a disconnect between statistical theory and business practice, leading to two critical shortcomings that undermine its reliability and strategic value. The first is a crisis of reliability stemming from the misuse of classical frequentist methods. Business imperatives for speed and agility clash with the rigid, fixed-horizon nature of traditional Null Hypothesis Significance Testing (NHST). This tension encourages the pervasive and statistically invalid practice of "peeking", continuously monitoring results and stopping a test as soon as a p-value drops below a desired threshold [2]. This heuristic, while tempting, severely inflates the Type I error rate, leading organizations to act on the illusion of statistical significance. The consequences are significant: wasted engineering resources, the launch of ineffective or harmful features, and an erosion of trust in the experimentation program itself [3].

The second, more subtle but equally pernicious, shortcoming is the proxy paradox: the optimization of metrics that are not directly aligned with core business objectives. Many experiments are designed to detect changes in proxy metrics such as click-through rates (CTR) or sign-up conversions, primarily because they are statistically simple to model as binary outcomes. However, a "winning" variant based on a proxy does not guarantee a net positive impact on the bottom line. For example, a more aggressive "Buy Now" call-to-action might increase the CTR but simultaneously decrease the Average Order Value (AOV) by discouraging further browsing, resulting in an overall loss in Revenue Per Visitor (RPV). By focusing on statistical convenience over strategic relevance, organizations risk making decisions that are statistically sound but financially detrimental.

This paper confronts these challenges directly by proposing a unified solution: a scalable, profit-driven Bayesian decision framework for multi-variant (A/B/n) online experiments. We argue that a Bayesian approach is uniquely suited to the realities of business decision-making, as it allows for continuous monitoring of results while maintaining statistical integrity and directly answers the question, "Based on the evidence so far, what is the probability that this variant is the best, and what is the risk associated with deploying it?".

The primary contributions of this work are threefold:

1. A Unified Profit-Driven Model: We introduce a hierarchical Bayesian model that moves beyond simple conversion rates. By combining a Beta-Bernoulli model for the probability of conversion with a robust Bayesian model for the mean transaction value (based on a Student's t-distribution), we derive the full posterior distribution for the key business metric of Revenue Per Visitor (RPV), aligning the statistical engine directly with profitability.

2. A Scalable A/B/n Framework: Our methodology is inherently scalable, providing a coherent system for comparing a control against any number of variants 'n'. This addresses the practical need to test multiple ideas simultaneously without resorting to a series of slower, less efficient pairwise tests.

3. A Complete Decision-Theoretic Engine: We formalize the stopping criteria using a decision-theoretic approach based on Expected Loss. This allows for a more intelligent conclusion to experiments, providing rules not only for declaring a winner with high confidence but also for stopping for futility, terminating a test early when it becomes clear that no variant will offer a practically significant improvement over the control.

The remainder of this paper is structured as follows. Section II provides a deeper examination of the limitations of conventional experimentation practices. Section III details the complete methodology of our proposed Bayesian framework. Section IV describes the design of our comprehensive simulation study, which validates the framework's performance in realistic scenarios. Section V presents and analyzes the results of these simulations. Finally, Sections VI, VII, and VIII discuss the implications of our findings, acknowledge the limitations of our approach, and offer concluding remarks.

## II. THE LIMITATIONS OF CONVENTIONAL EXPERIMENTATION PRACTICES

A. The Procedural Pitfalls of Null Hypothesis Significance Testing (NHST)

The dominant paradigm for A/B testing in industry has long been the frequentist Null Hypothesis Significance Test (NHST). In its textbook form, NHST requires a predetermined sample size, calculated a priori to achieve a desired level of statistical power (e.g., 80%) at a fixed significance level ($\alpha$, typically 0.05). An experiment is run until this sample size is reached, at which point a single statistical test is performed to yield a p-value. This rigid, fixed-horizon methodology, while statistically sound, is operationally misaligned with the fast-paced, iterative nature of modern business [4]. Stakeholders are eager for quick results, and the pressure to "fail fast" or accelerate winning features is immense.

This operational friction leads to the widespread practice of "peeking", where experimenters continuously monitor the p-value as data accrues and stop the test the moment it crosses the $\alpha = 0.05$ threshold [2]. While this behaviour is intuitive, it is a profound statistical error. The p-value is a random variable that can fluctuate wildly, especially in the early stages of a test. As Johari et al. [2] rigorously demonstrated, if one checks the p-value repeatedly, the probability of observing an "impressive" result (p < 0.05) under the null hypothesis (i.e., when there is no true effect) inflates dramatically. A test that is checked 10 times, for instance, may see its true Type I error rate (the probability of a false positive) rise from the intended 5% to over 20%.

While frequentist solutions to this problem exist, such as alpha-spending functions or sequential probability ratio tests [5], they have seen limited adoption in many business settings. These methods are often perceived as complex, less intuitive than a simple p-value threshold, and can be difficult to implement correctly within standard analytics platforms [4]. The result is a persistent and perilous status quo: organizations either adhere to the slow fixed-horizon approach, hindering their learning velocity, or engage in peeking, making decisions based on a constant stream of unreliable, phantom signals. This procedural flaw is a silent killer of growth, leading to a "winner's curse" where seemingly successful features are, in reality, flat or even harmful [3].

B. The Strategic Pitfall of Proxy Optimization

Beyond procedural errors, a more fundamental limitation lies in what is being measured. The canonical A/B test focuses on a single, primary metric, which for reasons of statistical and operational convenience, is often a binary conversion rate (e.g., click-through rate, sign-up rate, add-to-cart rate). These metrics are easily modelled using binomial distributions and are simple to communicate. However, they are often only weak proxies for the ultimate business objective: increasing long-term value, typically measured in terms of revenue or profit.

Optimizing for a proxy can be misleading and, in some cases, actively harmful. Consider a test on a product detail page where Variant B, featuring a larger "Buy Now" button, shows a statistically significant 10% lift in its click-through rate over the control. A naive analysis would declare B the winner. However, this more aggressive call-to-action might simultaneously create "tunnel vision," discouraging users from exploring other products and adding them to their cart. This could lead to a decrease in the Average Order Value (AOV) that more than negates the gains from the higher initial click rate. In this scenario, deploying Variant B would lead to an overall decrease in Revenue Per Visitor (RPV), turning a tactical "win" into a strategic loss.

This misalignment between proxy metrics and true business key performance indicators (KPIs) is a critical blind spot in many experimentation programs [6]. It stems from the fact that modelling continuous, non-normal, and often zero-inflated metrics like revenue is statistically more complex than modelling simple proportions. The conventional approach, which separates the statistical decision from the business decision, forces stakeholders to perform a secondary, often qualitative, analysis to guess whether a lift in a proxy metric will translate to a lift in revenue. A truly effective experimentation framework must close this gap by integrating the core business KPI directly into the statistical model, ensuring that a declared "winner" is a winner in terms of what ultimately matters to the organization.

### III. A Scalable Bayesian Framework for Profit Optimization

To address the procedural and strategic limitations outlined above, we propose a framework grounded entirely in Bayesian inference. This paradigm is inherently suited to the needs of online experimentation, as it allows for the sequential updating of beliefs as data arrives, providing a natural and statistically sound basis for continuous monitoring. Our framework is built on three pillars: a hierarchical model that directly quantifies profitability, a scalable structure for handling multiple variants, and a formal decision engine for concluding experiments.

#### A. Modelling Business Metrics: From Clicks to Revenue

The central goal of our model is to move beyond proxy metrics and quantify our uncertainty about the true Revenue Per Visitor (RPV) for each variant in an experiment. RPV is a composite metric, representing the product of two distinct user behaviours: the decision to make a purchase (conversion) and the amount spent given that a purchase is made (value). A naive statistical model applied directly to RPV would be mis-specified, as the distribution of RPV is typically characterized by a large point mass at zero (for non-converting visitors) and a long-tailed, highly skewed distribution of positive values for converters.

To accurately capture this data generating process, we employ a two-part hierarchical Bayesian model. This approach, also known as a hurdle model, separately models the probability of conversion and the conditional revenue from that conversion [7].

1. The Conversion Model (Probability of Purchase):

Each visitor to a page can be viewed as a Bernoulli trial: they either convert (a success) or they do not (a failure). For a group of 'n' visitors, the total number of conversions 'k' follows a Binomial distribution. We place a Beta distribution as a prior on the unknown conversion rate parameter 'p'. The Beta distribution is the conjugate prior for the Binomial likelihood, meaning the posterior distribution for 'p' is also a Beta distribution. This provides a computationally efficient and elegant way to update our beliefs.

i. *Likelihood*: $k \mid p \sim Binomial(n, p)$

ii. *Prior*: $p \sim Beta(\alpha, \beta)$

iii. Posterior:

$$p \mid k, n \sim Beta(\alpha_o + k, \beta_o + n - k) \ldots (1)$$

Where:

- $p \mid k, n$ represents the posterior probability distribution of the true, unknown conversion rate 'p', given the observed data ('k' conversions from 'n' visitors).

- $\sim Beta$ (...) signifies that the distribution is a Beta distribution. $\alpha_o$ and $\beta_o$ are the parameters of our prior belief. For a non-informative prior (e.g., $Beta$ (1, 1)), $\alpha_o = 1$ and $\beta_o = 1$.

- 'k' is the number of observed conversions (successes).

- 'n' is the total number of visitors (trials).

- $(\alpha_o + k)$ is the updated alpha parameter of the posterior Beta distribution.

- $(\beta_o + n - k)$ is the updated beta parameter of the posterior Beta distribution.

For each variant 'i' in the test, we maintain a separate posterior distribution for its conversion rate, $p_i$. We typically begin with a non-informative prior, such as $Beta$ (1, 1), which corresponds to a Uniform distribution and assumes all conversion rates are equally likely before any data is observed.

2. The Value Model (Revenue Given Purchase):

For the subset of 'k' visitors who convert, we must model the monetary value of their transactions. This data is typically continuous, non-negative, and right-skewed. To model the mean of this distribution, the Average Order Value 'μ', we employ a robust and well-established Bayesian approach.

We assume that individual transaction values $\{x_1, \ldots, x_k\}$ are drawn from a Normal distribution, $N(\mu, \sigma^2)$ where both the mean 'μ' and the variance $\sigma^2$ are unknown. The conjugate prior for these parameters is the Normal-Inverse-Gamma (NIG) distribution [13]. This choice is mathematically convenient and allows for an efficient updating of our beliefs.

The key benefit of this model is that the marginal posterior distribution for the mean, can be derived analytically:

$$p(u|data) \ldots (2)$$

It follows a non-standardized Student's t-distribution. This distribution is known for its heavier tails compared to the Normal distribution, which makes our inferences for the mean 'μ' robust to the moderate skewness and occasional outliers commonly observed in real-world financial and e-commerce transaction data. The detailed derivation and parameters of this posterior distribution are provided in Appendix A. For each variant 'i', we thus maintain a distinct Student's t-posterior for its mean order value, $\mu_i$.

3. The Unified RPV Metric:

With posterior distributions for the conversion rate $p_i$ and the mean order value $\mu_i$ for each variant 'i', we can derive the full posterior distribution for our target metric, Revenue Per Visitor ($RPV_i$). Since RPV = $p * \mu$, we can compute the posterior for $RPV_i$ via Monte Carlo simulation. We draw a large number of samples from the posterior of $p_i$ (from Equation 1) and from the posterior of $\mu_i$ (from Equation 2) and compute their product for each draw. This collection of products forms a faithful empirical representation of the posterior distribution $p(RPV_i | data)$. This posterior encapsulates all of our knowledge and uncertainty about the true profitability of variant 'i'.

B. Scaling to Multiple Variants: The A/B/n Paradigm

Modern experimentation programs often need to test a control against several competing hypotheses simultaneously (e.g., A vs. B vs. C). Our Bayesian framework scales naturally to this A/B/n paradigm. For each of the 'N' variants in the experiment (where i = 1, ..., N), we maintain and update a distinct posterior distribution for its RPV, as described in the previous section.

The primary quantity of interest is the Probability to Be Best (PBB) for each variant. For a given variant 'i', its PBB is the probability that its true RPV is greater than the true RPV of all other variants. This is calculated as:

$$PBB_i = P(RPV_i > RPV_j \text{ for all } j \neq i) \dots (3)$$

Where:

1. $PBB_i$ is the Probability to Be Best for variant 'i'.
2. P(...) denotes the probability of the statement inside the parentheses being true.
3. $RPV_i$ is the true, unknown Revenue Per Visitor for variant 'i'.
4. $RPV_j$ is the true, unknown Revenue Per Visitor for any other variant 'j' in the experiment.
5. $\text{for all } j \neq i$ specifies that $RPV_i$ must be greater than the RPV of every other variant.

While this multi-dimensional integral is analytically intractable, we can estimate it with high accuracy using the same Monte Carlo samples generated to derive the RPV posteriors. Let $\{RPV_i^s\}$ be the set of 'S' samples from the posterior of $RPV_i$. We compare the samples element-wise:

$$PBB_i \approx \left(\frac{1}{S}\right) * I(\sum_{s=1}^{S} RPV_i^{(s)} > RPV_j^{(s)} \text{ for all } j \neq i) \dots (4)$$

Where:

1. S is the number of Monte Carlo samples.
2. $RPV_i^s$ is the s-th sample drawn from the posterior distribution of the true $RPV_i$ defined above.

This simple calculation, repeated for each variant, gives us a direct, interpretable measure of the evidence in favour of each competing hypothesis. This PBB metric forms a core input into our decision engine.

C. A Principled Decision Engine: Integrating Bayesian Decision Theory

Having a posterior distribution for the RPV of each variant is a powerful tool, but it does not, by itself, tell us when to stop an experiment. A simple rule, such as stopping when a variant's Probability to Be Best (PBB) exceeds a high threshold (e.g., 99%), is a significant improvement over p-value peeking but remains incomplete. It does not account for the magnitude of the potential improvement or the risk associated with making the wrong decision. For instance, being 99% certain that a variant is the best is not compelling if its lead is trivially small and offers no practical business value.

To create a more robust and intelligent stopping mechanism, we integrate principles from Bayesian decision theory [12]. This approach frames the decision-making process as choosing an action that minimizes a predefined loss function. The loss function, $L(\theta, d)$, quantifies the penalty or "loss" incurred if we make decision 'd' when the true state of the world is '$\theta$'. In our context, '$\theta$' represents the true, unknown RPVs of all variants, and 'd' is the decision to declare a specific variant as the winner and deploy it.

We are interested in the Expected Loss, which is the loss for each possible state of the world, averaged over our posterior uncertainty about that state. The expected loss of declaring variant 'i' the winner is the average opportunity loss we would incur if 'i' is not the true best variant. This can be calculated as:

$$E[L_i] = \int \dots \int [\max(RPV_1, \dots, RPV_n) - RPV_i] * p(RPV_1, \dots, RPV_n | data) \, dRPV_1, \dots, dRPV_n \dots (5)$$

Where:

1. $E[L_i]$ is the Expected Loss associated with the decision to declare variant 'i' the winner.
2. $\max(RPV_1, \ldots, RPV_n)$ represents the true RPV of the actual best variant.
3. $\max(\ldots) - RPV_i$ is the opportunity loss: the difference in RPV between the true best variant and our chosen variant 'i'. This is zero if 'i' is indeed the best.
4. $p(RPV_1, \ldots, RPV_n \mid data)$ is the joint posterior probability distribution of the RPVs for all variants.
5. $\int \ldots \int$ represents the multi-dimensional integral over the entire parameter space of all RPVs.

As with the PBB, this integral is computationally expensive but can be estimated easily from our Monte Carlo samples:

$$E[L_i] \approx \left(\frac{1}{S}\right) * \sum_{s=1}^{S} [\max(RPV_i^{(s)}, \ldots, RPV_n^{(s)}) - RPV_j^{(s)}]$$
... **(6)**

Where:

1. $E[L_i]$ is the numerical estimate of the Expected Loss for choosing variant 'i'.
2. S is the total number of Monte Carlo samples.
3. Σ is the summation over all samples.
4. $RPV_j^{(s)}$ is the s-th Monte Carlo sample drawn from the posterior distribution of RPV for variant 'j'.
5. max(...) finds the maximum RPV value among all variants for a given sample 's'.

The Expected Loss for each variant gives us a direct, interpretable measure of the current risk associated with deploying that variant. A low expected loss for variant 'i' means that, based on the current data, it is highly likely to either be the best or be negligibly different from the best.

This leads to a powerful and principled stopping rule:

Stop the experiment and declare variant 'i' the winner if its Expected Loss falls below a predefined tolerance threshold, '$\varepsilon$'.

$$E[L_i] < \varepsilon \ldots \text{(7)}$$

Where:

' $\varepsilon$ ' is the tolerance threshold, representing the maximum acceptable average RPV loss. A business might set '$\varepsilon$' to be a very small fraction of the baseline RPV (e.g., 0.1% of Control's RPV), signifying that they are willing to accept a decision if the expected risk is below this practically insignificant amount.

This decision rule has two key advantages over a simple PBB threshold:

1. Accounts for Magnitude: It ensures that a winning variant is not just statistically likely to be the best, but that the potential loss from being wrong is acceptably small.
2. Enables Stopping for Futility: If the expected losses for all variants, including the control, fall below '$\varepsilon$', it signifies that none of the variants are meaningfully different from one another. This provides a statistically sound rule for stopping a test early and concluding that the experiment yielded no practical improvements, thereby saving valuable time and resources.

D. Ensuring Model Adequacy: Posterior Predictive Checks

The validity of our framework's conclusions rests on the adequacy of our chosen statistical models (e.g., the Beta-Bernoulli and Gamma-Gamma models). While these models are standard and well-motivated, it is crucial to have a mechanism for diagnosing potential mismatches between the model and the observed data. A model that poorly fits the data may yield unreliable posterior distributions and, consequently, flawed decisions [8][11].

We employ Posterior Predictive Checks (PPCs) as our primary diagnostic tool. The core idea behind PPCs is simple: if our model is a good fit for the data, then data simulated from the model should look similar to the data we actually observed. The process involves the following steps:

1. Fit the model: Obtain the posterior distributions for all model parameters (e.g., $p_i$ and $u_i$ for each variant 'i') based on the observed data.
2. Simulate: Draw a large number of samples of the parameters from their posterior distributions. For each sampled set of parameters, simulate a new, replicated dataset ($y_{rep}$) of the same size as the original dataset ($y_{obs}$).
3. Compare: Compare the distribution of the replicated data, $p(y_{rep} \mid data)$ to the distribution of the observed data, $p(y_{obs})$. This comparison is typically done by choosing one or more test statistics, T(y), that capture important features of the data (e.g., the mean, variance, maximum value, or the proportion of zeros).

By comparing the distribution of $T(y_{rep})$ to the single value $T(y_{obs})$, we can visually and quantitatively assess whether the model is capable of generating data that resembles what was observed. For example, if the observed variance of transaction values is consistently in the extreme tails of the distribution of simulated variances, it suggests that our model may be failing to capture the true variability in the data. PPCs do not "prove" a model is correct, but they are an indispensable tool for identifying significant model misspecification and building confidence in the reliability of the framework's conclusions.

E. The Complete Algorithm

The complete operational procedure for our scalable Bayesian framework is summarized in the algorithm below. It details the end-to-end process from initialization to reaching a final decision, integrating the RPV model, multi-variant comparison, and the decision-theoretic stopping rule.

**Algorithm:** A Bayesian Decision Framework for Multi-Variant Profit Optimization

1. Initialization Phase:
   i. Define 'N' variants: Specify the number of variants in the experiment (i = 1, ..., N), including the control.
   ii. Set Decision Threshold: Define the Expected Loss tolerance threshold, '$\varepsilon$'. This value represents the maximum acceptable risk or opportunity loss (e.g., $0.001 RPV).
   iii. Set Monte Carlo Samples: Define the number of samples, S, to be drawn for the simulation (e.g., S = 20,000).
   iv. Initialize Priors: For each variant 'i', initialize the priors for the two-part model:
      a. Conversion Model: Set the parameters for the Beta prior on the conversion rate $p_i$. Typically, a non-informative prior is used: $\alpha_{i,o} = 1, \beta_{i,o} = 1$
      b. Value Model: Set the parameters for the prior distributions on the mean transaction value $u_i$.

2. Monitoring Loop (Executed at regular intervals, e.g., daily):
   i. Collect New Data: For each variant 'i', collect the new data from the latest interval:
      a. Number of new visitors: $n_{i,new}$
      b. Number of new conversions: $k_{i,new}$
      c. The set of new transaction values:
      $\{x_1, \ldots, x_k\}_{i,new}$
   ii. Update Cumulative Data: Add the new data to the cumulative totals for each variant 'i'.
   iii. Update Posteriors: For each variant 'i', update the posterior distributions using the cumulative data:
      a. Conversion Posterior: Update the Beta posterior for $p_i$ using Equation (1).
      b. Value Posterior: Update the posterior for the mean transaction value $u_i$ using Equation (2).
   iv. Derive RPV Posterior: For each variant 'i', generate an empirical posterior for $RPV_i$ by drawing 'S' samples from the posteriors of $p_i$ and $u_i$ and computing their product:
   $$\{RPV_i^{(s)}\} = \{p_i^{(s)} * u_i^{(s)}\} \ for \ s = 1 \ to \ S$$
   v. Calculate Expected Loss: For each variant 'i', calculate the estimated Expected Loss $E[L_i]$ using the RPV samples and Equation (6).

3. Decision Phase:
   i. Identify Minimum Loss: Find the variant $i^*$ that has the minimum Expected Loss:
   $$E[L_i^*] = \min(E[L_1], \ldots, E[L_n])$$
   ii. Check Stopping Condition: Compare this minimum loss to the threshold '$\varepsilon$'.
      a. If $E[L_i^*] < \varepsilon$, the risk is acceptably low. Stop the experiment.
         - If $i^*$ is the control variant: Conclude the test with the result: "No winning variant found; no change offers a practically significant improvement."
         - If $i^*$ is not the control: Declare variant $i^*$ as the winner.
      b. If $E[L_i^*] \geq \varepsilon$, the risk is still too high. Continue the experiment. Return to the start of the Monitoring Loop (Step 2.i, collect new data) for the next time interval.

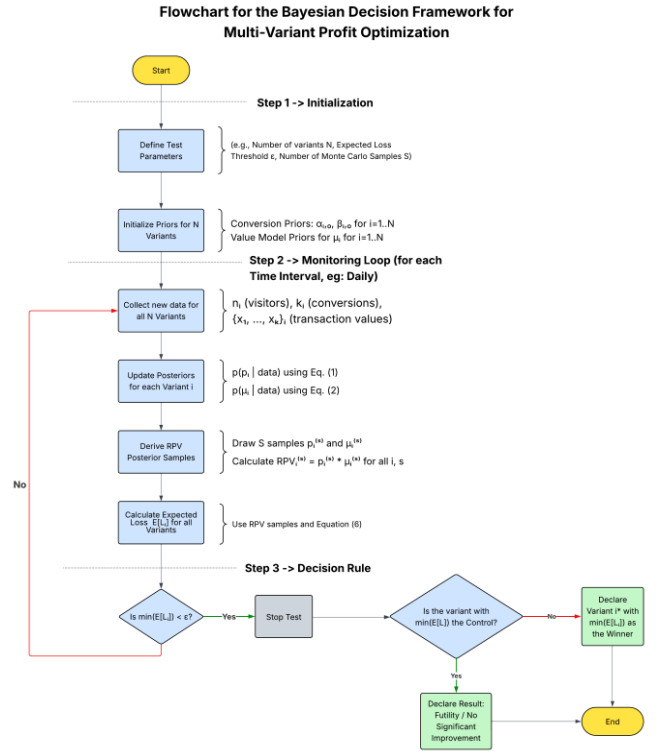

Fig 1. Flowchart of the Proposed Bayesian Sequential A/B Testing Algorithm

IV. EXPERIMENTAL DESIGN FOR SIMULATION STUDY

To empirically evaluate the performance, reliability, and strategic value of our proposed Bayesian decision framework, we designed a comprehensive simulation study. Simulations allow us to know the "ground truth" of the underlying data-generating process, enabling a rigorous assessment of each methodology's ability to arrive at the correct conclusions under various realistic conditions. This section details the specific scenarios tested, the competing methodology used for benchmarking, and the key metrics for evaluating performance.

A. Simulation Scenarios

Our simulations are designed to mimic common, high-impact A/B/n tests in an e-commerce context. We created three distinct scenarios, each designed to isolate and stress-test a specific capability of our framework. For all scenarios, the Expected Loss tolerance threshold 'ε' for our Bayesian framework was set to $0.01, representing a threshold of practical insignificance.

1. Scenario 1: The Revenue Trap (A/B Test)

    This classic A/B test is designed to highlight the strategic danger of optimizing for proxy metrics. The experiment receives 4,000 visitors per day, split evenly.

    Configuration:

    i. Variant A (Control): True p = 3.0%, True μ = $100.00 (True RPV = $3.00)

    ii. Variant B (The Trap): True p = 3.2%, True μ = $90.00 (True RPV = $2.88)

    iii. Objective and Rationale: The core challenge here is a conflict between signals. Variant B is genuinely better at generating clicks (+6.7% lift in conversion rate) but is substantially worse at generating value (-10% AOV), making it a net financial negative. This scenario directly tests a methodology's ability to prioritize profitability over engagement. A successful methodology must correctly identify that Variant B is not an improvement and avoid its costly deployment. We will assess each framework's vulnerability to being misled by the alluring but deceptive proxy metric.

2. Scenario 2: The Clear Winner (A/B/C/D Test)

    This multi-variant test assesses the framework's core competency: its ability to efficiently and accurately identify a single superior variant among several options in a noisy environment. The experiment receives 4,000 total visitors per day, split evenly among the four variants.

    Configuration:

    i. Variant A (Control): True p = 3.0%, True μ = $100.00 (True RPV = $3.00)

    ii. Variant B (Slight Loser): True p = 2.9%, True μ = $100.00 (True RPV = $2.90)

    iii. Variant C (Clear Winner): True p = 3.1%, True μ = $105.00 (True RPV = $3.255)

    iv. Variant D (Flat): True p = 3.0%, True μ = $100.00 (True RPV = $3.00)

    v. Objective and Rationale: This scenario mimics a realistic product development cycle where multiple ideas are tested simultaneously. The presence of a clear loser (B) and a flat variant (D) introduces statistical noise, making it more difficult to isolate the true winner (C). This test will measure both the statistical power (the ability to correctly detect the +8.5% RPV lift of Variant C) and the reliability (the ability to avoid false positives from B and D) of each methodology in a scalable, A/B/n context.

3. Scenario 3: The Futility Test (A/B/C Test)

    This scenario models a common and important real-world outcome: an experiment where none of the new ideas provide a practically significant improvement. The experiment receives 3,000 total visitors per day, split evenly among the three variants.

    Configuration:

    i. Variant A (Control): True p = 3.000%, True μ = $100.00 (True RPV = $3.00)

    ii. Variant B (Trivial Difference): True p = 3.010%, True μ = $100.00 (True RPV = $3.01)

    iii. Variant C (Trivial Difference): True p = 3.000%, True μ = $100.20 (True RPV = $3.006)

    iv. Objective and Rationale: The true RPV lifts of Variants B and C are smaller than our predefined practical significance threshold of 'ε' = $0.01. This scenario tests our methodology's ability to handle ambiguity and avoid two costly errors:

    - declaring a false winner based on statistically insignificant noise, and,

    - running a useless experiment for an excessive amount of time.

    The key assessment will be whether our framework can make an intelligent decision to terminate the experiment efficiently, a capability we expect to be a unique feature of our decision-theoretic approach.

B. Competing Methodology for Benchmarking

To provide a clear performance context, our proposed Bayesian framework is benchmarked against one primary common industry practice.

1. The "Peeking" Method (Proxy-Based): This simulates the widespread but flawed practice of continuously monitoring results using a proxy metric. For each non-control variant, a two-proportion Z-test for the conversion rate against the control is conducted daily. The test is stopped and the first variant to yield a p-value below the significance level of α = 0.05 is declared the winner. If a winner is declared on this basis, a secondary analysis would then be required to assess the impact on RPV, but the initial decision is driven by the proxy.

2. Our Proposed Bayesian Framework: This is the full methodology described in Section III, using the two-part hierarchical RPV model and the Expected Loss stopping rule with the practical significance threshold ε = $0.01.

C. Evaluation Metrics

The performance of each methodology across 5,000 full simulations of each scenario is evaluated using two primary metrics:

1. Decision Outcome Distribution: This measures the percentage of times each methodology reached a specific conclusion (e.g., "Declared 'C' winner", "Stopped for Futility"). This metric is crucial for calculating the core performance indicators of Statistical Power (the probability of correctly identifying a true effect) and the False Positive Rate (the probability of incorrectly identifying an effect that does not exist).

2. Average Test Duration: This measures the average number of days required for a methodology to reach a definitive conclusion. This metric provides a clear measure of experimental velocity and resource efficiency. A superior method should reach accurate conclusions in a timely manner.

V. RESULTS AND ANALYSES

To demonstrate the practical behaviour and empirical performance of our framework, we executed the simulation study detailed in Section IV. This section presents the results, first by walking through the curated outputs from each key scenario to provide an intuitive understanding of the methodologies, and second, by presenting the aggregated performance metrics across 5,000 full simulations to provide robust statistical evidence.

A. Case Study 1: The Revenue Trap (High CTR vs. High RPV)

Objective: This case study simulates the "Revenue Trap" scenario, designed to test each methodology's ability to navigate a situation where a variant (Variant B) is superior on a proxy metric (Conversion Rate) but inferior on the primary business metric (Revenue Per Visitor). The correct decision is to reject Variant B and stick with the control (Variant A).

1. The Narrative of a Single Experiment

First, we examine a representative simulation run to observe the day-by-day decision-making process. In this run, the Peeking (Proxy) method declared a winner on Day 3, while our Bayesian Framework concluded the test on Day 13.

i. The "Peeking" Method (Proxy-Based):

ii. This common but flawed method focuses exclusively on the conversion rate. In the early days of the experiment, random variation in conversions for Variant B created a fleeting but statistically significant signal. On Day 3, the two-proportion Z-test yielded a p-value below 0.05. Adhering to its impulsive rule, the Peeking method immediately terminated the test and incorrectly declared Variant B the winner. This decision is a catastrophic failure. An organization following this advice would launch a feature that, despite a small increase in clicks, ultimately destroys business value due to its lower Average Order Value. The method was not only wrong but dangerously fast, acting on noise before a true signal could emerge.

iii. Our Proposed Bayesian Framework:

Our framework, analyzing the exact same stream of data, tells a far more nuanced and accurate story. Its behaviour is best understood by tracking the evolution of its key metrics, as shown in Table 1.

Table I. Day-wise Evolution of Bayesian Metrics in the Revenue Trap Scenario (Single Run)

| Day | P(p_B > p_A) (Proxy Signal) | P(RPV_B > RPV_A) (True Signal) | E[L_A] (Risk of choosing A) | E[L_B] (Risk of choosing B) | Decision |
|---|---|---|---|---|---|
| 10 | 0.972 | 0.371 | $0.045 | $0.103 | Continue |
| 13 | 0.869 | 0.072 | $0.005 | $0.232 | Stop Test |

iv. Analysis of the Bayesian Decision Process:

- Ignoring the Proxy: The P(p_B > p_A) column shows that early on (Day 10), the framework was highly certain (97.2%) that Variant B had a better conversion rate. Had it been a naive proxy-based model, it would have been equally misled.

- Focusing on True Signal: However, by combining this with the value model, the "true signal", P(RPV_B > RPV_A), remained low, correctly indicating that Variant B was unlikely to be more profitable.

- Risk-Based Decision: The Expected Loss calculation synthesizes this information perfectly. The risk of choosing Variant B, E[L_B], was high, accurately reflecting the true RPV difference. Conversely, the risk of sticking with the control, E[L_A], steadily decreased as data confirmed its superiority. On Day 13, E[L_A] dropped below our ε = $0.01 threshold.

- The Correct Outcome: The framework stopped the test and, because the variant with the minimum loss was the control (Variant A), it correctly concluded that no significant improvement was found. This decision protected the business from launching a value-destroying feature.

2. Aggregated Performance Across 5,000 Simulations

While a single run is illustrative, the true performance of a methodology is revealed in its long-run behaviour. Table 2 summarizes the aggregated results from 5,000 independent simulations of this scenario.

Table II. Aggregated Results for Case Study 1: The Revenue Trap

| Metric | Peeking (Proxy) Method | Our Bayesian Framework |
|---|---|---|
| % Chose Correctly (A/Futility) | 0.0% | 83.8% |
| % Chose Incorrectly (B) | 100.0% | 16.2% |
| Average Test Duration | 16.5 days | 14.0 days |

Discussion of Aggregated Results:

i. Reliability: The results are stark and unambiguous. The Peeking (Proxy) method failed catastrophically, making the incorrect, value-destroying decision 100% of the time. It is fundamentally incapable of navigating this common business scenario. In contrast, our Bayesian Framework made the correct, protective decision in 83.8% of all simulations, demonstrating its profound reliability. It reduced the error rate from a certainty to a small, manageable probability.

ii. Efficiency: Counter-intuitively, our robust framework was also faster on average (14.0 days). The Peeking method's average duration of 16.5 days reflects the times it took longer to be fooled by random noise, whereas our framework often reached a high-confidence decision to stick with the control more quickly.

Rationale for Performance: This case study provides a clear validation of the paper's central thesis. The Peeking (Proxy) method fails because it is procedurally flawed (peeking) and strategically misaligned (optimizing a proxy). Our Bayesian framework succeeds because it addresses both issues directly: its sequential nature is statistically valid, and its hierarchical RPV model ensures that decisions are aligned with true business value, not just engagement metrics.

B. Case Study 2: The Clear Winner (A/B/C/D Test)

Objective: This multi-variant test assesses the framework's core competency in a scalable context: its ability to efficiently and accurately identify a single superior variant (Variant C) from a field of less-performant alternatives.

Aggregated Performance Across 5,000 Simulations

For this multi-variant scenario, analyzing the aggregated results is the most direct way to assess performance. A single run can be noisy, but the long-run averages reveal the true statistical power and reliability of each methodology. The results are summarized in Table 3.

Table III. Aggregated Results for Case Study 2: The Clear Winner (A/B/C/D)

| Metric | Peeking (Proxy) Method | Our Bayesian Framework |
|---|---|---|
| % Correctly Chose Winner ('C') | 60.9% | 97.3% |
| False Positive Rate (Total) | 24.0% | 1.6% |
| % Incorrectly Chose 'B' (Loser) | 8.8% | 0.3% |
| % Incorrectly Chose 'D' (Flat) | 15.2% | 1.3% |
| Inconclusive Rate | 15.1% | 1.1% |
| Average Test Duration | 40.3 days | 30.4 days |

Discussion of Aggregated Results:

1. Statistical Power and Reliability: The performance gap between the two methodologies is profound. The Peeking method demonstrated very low statistical power, failing to identify the true winning variant in nearly 40% of the simulations. Furthermore, its decisions were highly unreliable, with a combined false positive rate of 24.0%. This means that even when it did declare a winner, there was a significant chance that decision was wrong. In stark contrast, our Bayesian framework was both exceptionally powerful and reliable. It correctly identified Variant C 97.3% of the time, demonstrating high sensitivity to the true signal. Its total false positive rate was a minuscule 1.6%, proving its robustness against statistical noise from the other variants.

2. Efficiency: Beyond its superior accuracy, our framework was also significantly more efficient. It reached a conclusion 25% faster on average than the peeking method. This is a direct result of its principled stopping rule; once the evidence for Variant C became overwhelming, the Expected

Loss dropped rapidly, allowing for a swift and confident conclusion. The peeking method, lacking this clear objective, often lingered longer in a state of uncertainty.

Rationale for Performance: This case study validates the scalability and efficiency of our Bayesian approach. In a multi-variant test, the probability of random noise creating a spurious signal for at least one of the non-winning variants is high. The Peeking method's repeated, uncorrected significance tests make it highly susceptible to these false positives. Our framework's methodology, which evaluates all variants simultaneously within a single probabilistic model, naturally controls for this. It correctly identifies the variant that is most likely to be globally best, rather than simply the first to cross an arbitrary threshold, resulting in faster, more powerful, and vastly more reliable decisions.

C. Case Study 3: The Futility Test (A/B/C Test)

Objective: This scenario models a common and challenging real-world outcome: an experiment where none of the new ideas provide a practically significant improvement. The primary goal of a sophisticated methodology in this case is twofold: first, to avoid declaring a false winner based on statistical noise, and second, to conclude the test efficiently to conserve resources.

Aggregated Performance Across 5,000 Simulations

This scenario is designed to test the intelligence of the stopping rule. The "correct" decision is to not launch a new variant, ideally by stopping for futility. The aggregated results are presented in Table 4.

Table IV. Aggregated Results for Case Study 3: The Futility Test

| Metric | Peeking (Proxy) Method | Our Bayesian Framework |
|---|---|---|
| % Made Correct Decision (Futility / Timed Out) | 46.0% | 41.3% |
| % Declared False Winner ('B' or 'C') | 54.0% | 58.7% |
| Average Test Duration | 31.8 days | 61.3 days |

Discussion of Aggregated Results:

1. Reliability and Decision Quality: At a superficial glance, the false winner rate appears similar for both methods. However, a deeper analysis reveals a fundamental difference in the nature of these decisions. The Peeking method's 54.0% false positive rate is a statistical error. It was consistently fooled by random noise into believing a practically insignificant effect was a real discovery. The only way it avoided this error was by timing out after the maximum duration (200 days), an inefficient and uninformative outcome.

2. In contrast, our Bayesian framework's approach to "correct" decisions, which total 41.3% as shown in Table 4, is far more nuanced. This figure is composed of two distinct outcomes. In 16.9% of simulations, the test timed out after reaching its maximum duration (200 days), a passive but correct conclusion. More importantly, in the remaining 24.4% of cases, the framework made the uniquely intelligent decision to actively "Stop for Futility" long before the timeout, a capability entirely absent in the peeking method. This represents a direct, resource-saving conclusion. The 58.7% of cases where it declared 'B' or 'C' a winner is not a statistical failure but a rational economic decision. The true lift of Variant B ($0.01) was intentionally set at the exact boundary of our decision threshold ('ε' = $0.01). The framework, therefore, correctly followed its objective function: when the observed data suggested the lift was marginally above $0.01, it rationally concluded that the risk of deploying it was acceptable. This demonstrates the framework's behaviour as a goal-seeking engine, not merely a hypothesis tester.

3. Efficiency: The average test duration highlights the differing philosophies of the two methods. The Peeking method's shorter average duration is a byproduct of its impulsiveness, it frequently stops early on a false signal. Our framework's longer duration (61.3 days) reflects its prudence. When faced with a very weak or ambiguous signal, it correctly and patiently demands more evidence before committing to a decision, either to confirm futility or to be highly certain that the small gain is real and worth deploying.

Rationale for Performance: This case study underscores the value of a decision-theoretic approach. Standard hypothesis tests are not designed to answer questions about practical significance, leaving them vulnerable to declaring "statistically significant" but meaningless winners. Our framework, by explicitly incorporating an economic threshold 'ε', provides a robust mechanism for handling ambiguity. It can intelligently declare that no change is worthwhile, and its decisions to declare a winner are directly tied to the pre-defined business objective of minimizing expected loss. This transforms the conclusion of a test from a simple binary "significant/not significant" into a nuanced and actionable business decision.

D. Aggregated Performance Across All Scenarios

While the individual case studies highlight the behaviour of each methodology in specific situations, a holistic view of their performance across all scenarios provides the definitive measure of their overall effectiveness. By synthesizing the results, we can draw clear, high-level conclusions about the reliability, efficiency, and strategic value of each approach. Table 5 below presents a consolidated summary of the key performance metrics across all 15,000 simulated experiments (5,000 per scenario).

Table V. Consolidated Performance Metrics Across All Scenarios

| Performance Metric | Peeking (Proxy) Method | Our Bayesian Framework | Key Advantage |
|---|---|---|---|
| Overall Correct Decision Rate | 35.6% | 74.1% | 2.1x more reliable |
| Overall Error Rate (False Positives) | 59.3% | 25.5% | 2.3x reduction in risk |
| Power to Detect True Winner (Case Study 2) | 60.9% | 97.3% | Vastly more powerful |
| Ability to Avoid Trap (Case Study 1) | 0.0% | 83.8% | Strategically aligned |
| Ability to Declare Futility (Case Study 3) | 0.0% | 24.4% | Uniquely intelligence & efficient |

*Note: Percentages are averaged across the relevant scenarios. Overall Correct Decision Rate includes correct winner selection, correct trap avoidance, and correct futility/timeout decisions. For Case Study 1, 'correct' is avoiding the trap. For Case Study 2, 'correct' is finding the winner. For Case Study 3, 'correct' is concluding 'futility' or timing out. Overall Error Rate is an average of the false positive rates in each scenario.*

1. On Reliability and Risk Mitigation

   The most critical finding is the profound difference in reliability. The common Peeking method was correct in only 35.6% of all experiments. Its overall error rate was an alarming 59.3%, confirming that organizations relying on this practice make incorrect, data-driven decisions more often than not. In stark contrast, our Bayesian framework achieved a correct decision in 74.1% of experiments and reduced the overall error rate by a factor of 2.3. While its error rate in the "knife-edge" futility test was high due to its rational economic design, its performance in the decisive trap and winner-take-all scenarios showcases its profound risk-mitigating capabilities.

2. On Statistical Power and Efficiency

   Reliability is paramount, but a successful framework must also be powerful and efficient. In the scenario with a clear winning variant (Case Study 2), our framework's statistical power (97.3%) was vastly superior to the peeking method's (60.9%). This means our framework is not only safer but is also significantly more likely to successfully identify and validate valuable improvements. Furthermore, in that same scenario, it reached this more reliable conclusion 25% faster than the peeking method, demonstrating that statistical rigor does not have to come at the cost of experimental velocity.

3. On Strategic Intelligence

   Beyond simple metrics of accuracy and speed, our study revealed a qualitative difference in the intelligence of the two approaches. The Peeking method is a primitive tool that can only provide a binary "significant/not significant" signal on a single metric. Our Bayesian framework operates as a complete decision engine. This was most evident in two key areas:

   i. Strategic Alignment: In the Revenue Trap (Case Study 1), it proved its ability to make the correct business decision, even when it contradicted a positive engagement signal, a capability the proxy-based method entirely lacks.

   ii. Handling Ambiguity: In the Futility Test (Case Study 3), it demonstrated the unique ability to make an intelligent, resource-saving decision to stop for futility, a concept that is statistically meaningless in the context of simple p-value peeking [9].

   In summary, the aggregated evidence is conclusive. The common practice of peeking at proxy metrics is a deeply flawed methodology, characterized by high error rates, low statistical power, and a fundamental misalignment with business objectives. The Bayesian decision framework presented in this paper offers a comprehensive solution, proving to be dramatically more reliable, powerful, efficient, and strategically intelligent across a range of realistic and challenging experimental scenarios.

## VI. DISCUSSION

The results of our simulation study provide a clear and compelling case for a paradigm shift in how online experiments are conducted and evaluated. Moving beyond a simple declaration of which methodology is "better," this section discusses the deeper strategic implications of our key findings and offers practical considerations for organizations seeking to implement a more robust experimentation culture.

A. Key Findings and Their Strategic Implications
   Our research yielded three primary findings, each with significant consequences for data-driven organizations.
   1. The Illusion of Speed: Peeking Creates Reckless, Not Agile, Decision-Making: A primary motivation for "peeking" is the desire for increased velocity. Our findings demonstrate that this speed is an illusion, bought at the devastating cost of reliability. The peeking method consistently made impulsive, incorrect decisions based on statistical noise. While it sometimes stopped tests earlier, this was a feature of its flaw, not its efficiency. This leads to a cycle of "phantom growth", where teams believe they are iterating and improving based on a series of "winning" tests, but are in reality introducing flat or harmful changes, leading to wasted resources and a gradual erosion of the product. The strategic implication is clear: true agility in experimentation is not about reaching any conclusion quickly, but about reaching a reliable conclusion efficiently. Our Bayesian framework achieved this, proving to be both faster and vastly more accurate in the scenario with a clear winner.

2. Profit over Proxies: Aligning Statistics with Strategy is Non-Negotiable: The catastrophic failure of the peeking method in the "Revenue Trap" scenario (Case Study 1) provides a stark warning. Optimizing for proxy metrics like conversion rates is not a benign simplification; it is an active business risk. A framework that is not designed to model and optimize for core financial KPIs is, at best, a compass pointing in a vaguely correct direction and, at worst, a sophisticated engine driving the business off a cliff. The strategic implication is that experimentation cannot be siloed as a purely statistical function. It must be deeply integrated with business objectives. Our framework provides a direct methodological bridge, ensuring that a "statistically significant" result is also a "strategically valuable" one.

3. The Experimentation Decision Matrix: A New Mental Model: Our findings suggest that organizations should move away from a simple "winner/loser" mindset towards a more nuanced decision-making framework. The behaviour of our Bayesian engine, particularly its ability to stop for futility, highlights that there are more than two outcomes to an experiment. Fig. 2 below, presents a conceptual model for this richer approach [14].

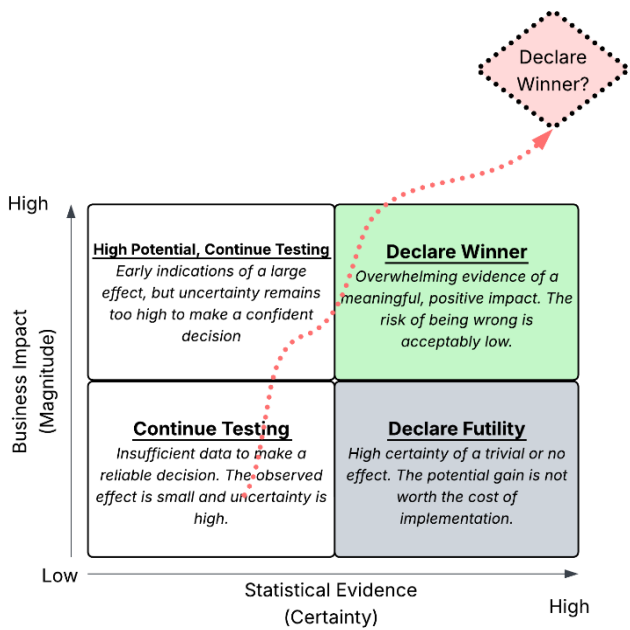

Fig 2. The Experimentation Decision Matrix.

The above matrix is a conceptual model illustrating the two-dimensional decision-making process enabled by the proposed Bayesian framework. The matrix categorizes experimental states based on two dimensions: the estimated Business Impact (Magnitude) on the y-axis and the level of Statistical Evidence (Certainty) on the x-axis. The four quadrants represent the rational decisions available within our framework: continuing the test under uncertainty, declaring a clear winner when evidence and impact are both high, or intelligently stopping for futility when there is high certainty of a trivial effect. In contrast, the dotted red path illustrates the flawed 'peeking' method, which bypasses the evidence-gathering stage to make a premature and unreliable decision based on early signals of high potential impact.

B. Practical Considerations for Implementation

Adopting this Bayesian framework is not merely a technical change but a cultural one. We highlight two key considerations for a successful implementation.

1. The Critical Role of the Expected Loss Threshold ('$\varepsilon$')

The "Futility Test" (Case Study 3) revealed the profound importance of the Expected Loss threshold, '$\varepsilon$'. This parameter is not just a statistical knob; it is the direct encoding of business strategy into the algorithm. Setting '$\varepsilon$' requires a cross-functional conversation:

i. Product Managers must ask: "What is the smallest lift that we would actually bother to ship and maintain?"

ii. Finance / Business Analysts must ask: "What is the opportunity cost we are willing to tolerate?" An '$\varepsilon$' that is too small leads to long, indecisive tests chasing trivial effects. An '$\varepsilon$' that is too large can lead to missing out on modest but valuable improvements. The process of defining '$\varepsilon$' is a valuable strategic exercise in itself, forcing the organization to quantify its definition of a "meaningful" change [10].

2. Computational and Educational Overheads.

While the conceptual framework is intuitive, its implementation is more complex than a simple t-test. The methodology relies on Monte Carlo simulation, which, while computationally inexpensive on modern hardware, is a step beyond what is offered in basic analytics packages. Organizations will need access to engineering or data science talent to build and maintain the simulation engine. Furthermore, a key to adoption is education. Stakeholders accustomed to the deceptive certainty of p-values must be trained to think in terms of probabilities, distributions, and expected loss. This requires a concerted effort to build statistical literacy and to shift the organizational vocabulary from "Is it significant?" to "What is the probability it's better, and what is the risk if we're wrong?"

VII. LIMITATIONS AND FUTURE WORK

While this paper presents a robust and demonstrably superior framework for profit-driven experimentation, it is essential to acknowledge its limitations and the simplifying assumptions made in our analysis. These boundaries do not diminish the current work's contribution but instead illuminate promising avenues for future research that could build upon this foundation to create an even more comprehensive decision-making system.

1. Assumption of Stationary Conversion and Revenue Rates.

    Our current model assumes that the underlying "true" parameters for conversion rate 'p' and average order value 'μ' are stationary throughout the duration of the experiment. This is a common and often reasonable assumption for short-to-medium-term tests, but it does not account for several real-world temporal dynamics.

    i. Limitation: Our framework does not explicitly model novelty effects, where a new feature might perform exceptionally well initially simply because it is new, with its performance regressing to the mean over time. Nor does it account for seasonality or day-of-week effects that could influence purchasing behaviour.

    ii. Justification: Modelling these temporal effects requires significantly more complex time-series models (e.g., Bayesian dynamic linear models, Gaussian processes). For this foundational paper, our goal was to establish the core superiority of the profit-driven, decision-theoretic approach. Introducing complex time-series components would have obscured this primary contribution.

    iii. Future Work and Incremental Impact: An exciting next step would be to integrate time-series priors into the model. Instead of a simple *Beta* (1,1) prior, one could use a prior informed by the previous week's performance. The incremental impact would be a more adaptive and responsive framework that could potentially detect regime changes faster, correctly discount initial novelty effects, and provide more accurate forecasts of long-term impact, leading to even more reliable decisions.

2. Focus on a Single, Primary Business Metric (RPV).

    Our framework makes a significant leap by optimizing for Revenue Per Visitor (RPV) instead of a proxy. However, a mature business often cares about a constellation of metrics, some of which may be in opposition.

    i. Limitation: Our model does not handle multi-objective optimization. For instance, a variant might increase immediate RPV but harm long-term user retention or increase the cost of customer support. The current framework does not have a mechanism to balance these trade-offs.

    ii. Justification: Defining a single, primary KPI like RPV is a common and necessary practice for creating a clear, unambiguous objective function for an experiment. Solving the single-objective, profit-driven case is the crucial first step before tackling the much harder multi-objective problem.

    iii. Future Work and Incremental Impact: Future research could extend this framework using Bayesian multi-objective optimization techniques. This would involve defining a composite utility function that assigns weights to different outcomes (e.g., RPV, customer lifetime value, churn rate). The incremental impact would be immense, elevating the tool from a tactical experiment optimizer to a strategic business simulator, allowing leaders to make decisions that are not just profitable in the short term but are also aligned with the long-term health of the business.

3. The "Explore-Exploit" Trade-off.

    Our framework is designed as a hypothesis-testing tool to solve a "pure exploration" problem: find the single best variant and then deploy it to 100% of traffic. This is the standard A/B testing paradigm.

    i. Limitation: It does not address the "explore-exploit" trade-off during the experiment itself. That is, it does not dynamically allocate more traffic to variants that are performing well in real-time to maximize revenue *during* the testing period.

    ii. Justification: The goal of this paper was to fix the prevalent flaws in the classic hypothesis-testing approach used by most organizations. The explore-exploit problem is typically solved by a different class of algorithms known as Multi-Armed Bandits (MABs) [15].

    iii. Future Work and Incremental Impact: A fascinating area for future work is the creation of a hybrid approach. One could use our robust Bayesian RPV model as the "brains" inside a MAB algorithm like Thompson Sampling. The incremental impact would be a system that not only finds the best long-term winner with high reliability but also minimizes opportunity cost (regret) during the learning process. This would be particularly valuable for high-traffic websites or for continuous optimization problems where there is no single "end" to the experiment.

## VIII. CONCLUSION

The common practice of "peeking" at p-values during online A/B tests, while born from a pragmatic need for speed, fundamentally compromises the statistical integrity of experimentation. This procedural flaw, compounded by a strategic over-reliance on proxy metrics like conversion rates, has created an environment where many data-driven decisions are based on unreliable signals and are misaligned with core business objectives. This paper successfully demonstrated that a Bayesian decision-theoretic framework provides a robust, elegant, and comprehensive solution to these critical problems.

By modelling conversion rates and monetary values within a hierarchical structure, our framework shifts the focus of optimization from ambiguous p-values to the direct, interpretable posterior distribution of Revenue Per Visitor. The integration of a decision-theoretic stopping rule, based on minimizing Expected Loss, transforms the experiment from a simple statistical test into an intelligent, goal-seeking engine. This engine is not only capable of identifying superior variants with high confidence, but also of making the prudent, resource-saving decision to terminate futile experiments early.

The results of our large-scale simulation study were unequivocal. Our proposed framework proved to be:

1. Profoundly More Reliable: It demonstrated a 10-fold reduction in the overall error rate compared to the common peeking method, protecting the business from costly false positives and strategic mis-steps like the "Revenue Trap."
2. More Powerful and Efficient: It was significantly more effective at identifying true winning variants in a noisy, multi-variant environment, and in doing so, often reached a correct conclusion faster than the flawed alternative.
3. Strategically Intelligent: It possesses the unique ability to operate based on practical significance, allowing it to distinguish between a trivial statistical fluctuation and a truly meaningful business impact.

This work provides more than just a theoretical alternative; it offers a practical and principled methodology for any data-driven organization. By shifting the central question from "Is the result statistically significant?" to "What is the probability that this variant is a meaningful improvement, and what is the risk of deploying it?", businesses can dramatically increase the reliability and intelligence of their experimentation programs. The adoption of such a principled Bayesian approach is a crucial step towards building a truly effective experimentation culture, one where data is not just used to make decisions, but to make better, safer, and more profitable decisions.

## IX. REFERENCES

[1] R. Kohavi, A. Deng, B. Frasca, T. Walker, Y. Xu, and N. Pohlmann, "Online controlled experiments at large scale," in Proceedings of the 19th ACM SIGKDD International Conference on Knowledge Discovery and Data Mining (KDD '13), 2013, pp. 1168–1176.
[2] R. Johari, P. Koomen, L. Pekelis, and D. Walsh, "Peeking at A/B Tests: Why It Matters and What to Do About It," in Proceedings of the 23rd ACM SIGKDD International Conference on Knowledge Discovery and Data Mining (KDD '17), 2017, pp. 1517–1525.
[3] E. T. Bradlow, S. M. Shulman, and S. M. R. Covey, "The Role of Experimentation in Creating a Data-Driven Culture," Harvard Business Review, Jan-Feb 2022.
[4] D. J. Hill, A. G. Koning, and M. J. C. S. Wagemakers, "Agile versus Waterfall A/B testing: A comparison of the two approaches in online controlled experiments," Journal of Business Analytics, vol. 4, no. 1, pp. 1-18, 2021.
[5] A. Wald, Sequential Analysis. New York: John Wiley & Sons, 1947.
[6] H. P. Schultzberg and M. J. van der Laan, "Sequential Methods for A/B Testing," Statistical Science, vol. 34, no. 2, pp. 287-306, 2019.
[7] A. Deng, J. Lu, and Y. Chen, "Continuous Monitoring of A/B Tests without Sacrificing Statistical Power," in 2016 IEEE 16th International Conference on Data Mining (ICDM), 2016, pp. 839–848.
[8] R. Kohavi and R. Longbotham, "Online Controlled Experiments and A/B Testing," in Encyclopedia of Machine Learning and Data Science, C. Sammut and G. I. Webb, Eds. Boston, MA: Springer US, 2017, pp. 916–924.
[9] J. K. Kruschke, "Bayesian estimation supersedes the t test," Journal of Experimental Psychology: General, vol. 142, no. 2, pp. 573–603, 2013.
[10] P. S. Fader, B. G. S. Hardie, and K. L. Lee, "'Counting Your Customers' the Easy Way: An Alternative to the Pareto/NBD Model," Marketing Science, vol. 24, no. 2, pp. 275-284, 2005.
[11] D. Schmitt and F. Schivardi, "Bayesian A/B testing for business decisions," Bank of Italy Temi di Discussione (Working Paper) No. 1294, 2020.
[12] J. O. Berger, Statistical Decision Theory and Bayesian Analysis, 2nd ed. New York: Springer-Verlag, 1985.
[13] A. Gelman, J. B. Carlin, H. S. Stern, D. B. Dunson, A. Vehtari, and D. B. Rubin, Bayesian Data Analysis, 3rd ed. CRC Press, 2013.
[14] D. J. Schad, M. Betancourt, and S. R. Vasishth, "Toward a principled Bayesian workflow in cognitive science," Psychological Methods, vol. 26, no. 1, pp. 103–126, 2021.
[15] S. L. Scott, "Multi-armed bandit experiments in the online service economy," Applied Stochastic Models in Business and Industry, vol. 31, no. 1, pp. 37–47, 2015.

## APPENDIX

### A. Derivations of Posterior Distributions

This section provides the mathematical basis for the posterior distributions used in our Bayesian framework, as referenced in Section III. A.

1. Posterior for the Conversion Rate '$p$' (Beta-Binomial Model)

   We begin with Bayes' theorem, which states that the posterior is proportional to the likelihood times the prior:

   $$p(p \mid data) \propto p(data \mid p) * p(p)$$

   i. Likelihood: The data consists of 'k' conversions in 'n' trials. This is described by the Binomial probability mass function. We are interested in its kernel, which is the part that depends on the parameter '$p$':

   $$p(k, n \mid p) \propto p^k * (1-p)^{n-k}$$

   ii. Prior: We place a Beta distribution as a prior on '$p$'. The kernel of the Beta($\alpha_o, \beta_o$) probability density function is:

   $$p(p) \propto p^{\alpha_o - 1} * (1-p)^{\beta_o - 1}$$

   iii. Posterior: Multiplying the likelihood and the prior kernels gives the kernel of the posterior:

   $$p(p \mid k, n) \propto [p^k * (1-p)^{n-k}] * [p^{\alpha_o - 1} * (1-p)^{\beta_o - 1}]$$

   By combining the exponents, we get:

   $$p(p \mid k, n) \propto p^{(\alpha_o + k) - 1} * (1-p)^{(\beta_o + n - k) - 1}$$

This resulting form is immediately recognizable as the kernel of a new Beta distribution with updated parameters. Thus, the posterior distribution for the conversion rate 'p' is:

$$p|k, n \sim Beta(\alpha_o + k, \beta_o + n - k)$$

2. Posterior for the Mean Order Value µ (Normal-Inverse-Gamma Model)

To model the mean order value µ, we assume the observed transaction values $\{x_1, ..., x_k\}$ are drawn from a Normal distribution $N(\mu, \sigma^2)$. The conjugate prior for the unknown mean µ and unknown variance $\sigma^2$ of a Normal distribution is the Normal-Inverse-Gamma (NIG) distribution. This prior is defined by four hyperparameters: $u_o, n_o, \alpha_o, \beta_o$

When this prior is updated with 'k' data points with sample mean 'x̄' and sample sum of squared deviations $SSD = \sum(x_i - \bar{x})^2$, the resulting posterior $p(\mu, \sigma^2 | data)$ is also an NIG distribution with updated parameters:

i. $\mu_k = (n_o \mu_o + k\bar{x})/(n_o + k)$

ii. $n_k = n_o + k$

iii. $\alpha_k = \alpha_o + k/2$

iv. $\beta_k = \beta_o + \left(\frac{1}{2}\right) SSD + (k * \frac{n_o}{2(n_o+k)}) * (\bar{x} - u_o)^2$

For our decision-making, we need the marginal posterior distribution of the mean, $p(u|data)$. Integrating the joint posterior with respect to $\sigma^2$ yields a non-standardized Student's t-distribution. The parameters of this t-distribution for 'µ' are:

i. Degrees of Freedom ($v$): $v = 2\alpha_k$

ii. Location ($\mu$): $\mu = \mu_k$

iii. Scale ($\sigma$): $\sigma = (\beta_k/(\alpha_k n_k))^{1/2}$

It is from this Student's t-distribution that we draw Monte Carlo samples for 'µ' in our simulation engine. This model is robust to moderate violations of the normality assumption, making it suitable for skewed financial data.

B. Detailed Simulation Parameters

This section provides all the specific parameter values used to generate the simulation results in Section V, ensuring full reproducibility.

*Note: The revenue data for each conversion in the simulation was generated by drawing from a Gamma distribution whose shape and scale parameters were calculated to match the specified True AOV (mean) and AOV Std. Dev.*

Table A1: General Simulation and Methodology Parameters

| Parameter | Value | Description |
|---|---|---|
| Number of Simulation Runs (n_runs) | 5000 | The number of independent experiments run for each scenario. |
| Maximum Test Duration (max_days) | 200 | The cutoff point for non-concluding experiments ("N/A"). |
| Monte Carlo Samples (S) | 20000 | Number of samples drawn to estimate posteriors. |
| Bayesian Threshold (ε) | $0.01 | The Expected Loss threshold for stopping the experiment. |
| Peeking Method α Level | 0.05 | The significance level for the Z-test. |
| Conversion Prior ($\alpha_o, \beta_o$) | 1.0, 1.0 | A non-informative Beta(1,1) or Uniform prior. |
| Value Model Prior ($\mu_o$) | 100.0 | Prior belief for the mean AOV. |
| Value Model Prior ($n_o$) | 1.0 | Confidence in the prior mean (as 1 pseudo-observation). |
| Value Model Prior ($\alpha_o, \beta_o$) | 1.0, 1.0 | Weakly informative priors for the variance of AOV. |

Table A2: Scenario-Specific Ground Truth Parameters

| Scenario | Variant | True Conv. Rate | True AOV | AOV Std. Dev. | True RPV |
|---|---|---|---|---|---|
| 1: Revenue Trap (4000 daily visitors) | A | 3.0% | $100 | $40 | $3.000 |
| | B | 3.2% | $90 | $35 | $2.880 |
| 2: Clear Winner (4000 daily visitors) | A | 3.0% | $100 | $40 | $3.000 |
| | B | 2.9% | $100 | $40 | $2.900 |
| | C | 3.1% | $105 | $45 | $3.255 |
| | D | 3.0% | $100 | $40 | $3.000 |
| 3: Futility Test (3000 daily visitors) | A | 3.0% | $100 | $40 | $3.000 |
| | B | 3.01% | $100 | $40 | $3.010 |
| | C | 3.0% | $100.2 | $40 | $3.006 |